\documentclass[natbib]{emulateapj}

\input epsf

\def\s2n{S^{\prime}/N}

\def\vv{{\bf V}}
\def\xx{{\bf x}}

\def\div{\nabla\cdot}

\begin{document}
\title{Confinement--Driven Spatial Variations in the Cosmic Ray Flux}
\author{Paolo Padoan\altaffilmark{1}  and John Scalo\altaffilmark{2}}
\altaffiltext{1}{Department of Physics, University of California, San Diego, 
CASS/UCSD 0424, 9500 Gilman Drive, La Jolla, CA 92093-0424; ppadoan@ucsd.edu}
\altaffiltext{2}{Department of Astronomy, University of Texas, Austin, TX 78712; parrot@astro.as.utexas.edu}

\begin{abstract}

Low--energy cosmic rays (CRs) are confined by self--generated MHD waves in the 
mostly neutral ISM. We show that the CR transport equation can 
be expressed as a continuity equation for the CR number density 
involving an effective convection velocity. Assuming balance between wave 
growth and ion--neutral damping, this equation gives a steady--state condition 
$n_{\rm cr} \propto n_{\rm i}^{1/2}$ up to a critical density for free streaming. 
This relation naturally accounts for the heretofore unexplained difference in CR 
ionization rates derived for dense diffuse clouds (McCall et al. 2003) and dark clouds, 
and predicts large spatial variations in the CR heating rate and pressure.

\end{abstract}


\section{Introduction}

Low--energy cosmic rays (CRs) 
are coupled to a number of important astrophysical processes.   
Most of the integrated interstellar CR 
number density, energy density, pressure and ionization rate are 
contributed by the low--energy part of the spectrum because of its 
steepness (see compilation in Antoni et al. 2004). 
\nocite{Antoni+04}
1--10 GeV CRs control the ionization fraction in the Earth's lower 
atmosphere, and so cloud formation and lightning production 
\citep[e.g.][]{Stozhkov03}. Tropospheric and stratospheric 
chemistry are affected, even at energies as low as $\sim 10$~MeV
\citep[e.g.][]{Crutzen+75}. In protostellar disks, ionization by  
ambient low--energy CRs may regulate disk turbulence 
and affect planet formation \citep[e.g.][]{Matsumura+Pudritz05}. On the scale 
of the interstellar medium, low--energy CRs of $\sim$100 MeV (Webber 1998) dominate 
the ionization fraction and heating of cool neutral gas, especially dark
UV--shielded molecular regions \citep{Goldsmith+Langer78}.
On larger scales, CR ``pressure'', most of which is
contributed by low--energy CRs, may help confine the Galactic disk and
drive the Parker instability \citep[see][]{Hanasz+Lesch00},
which may itself contribute to the formation of large condensations that
form stars, and even drive a Galactic dynamo \citep{Parker92}; the
CR pressure may also affect the hot coronal ISM
\citep{Schlickeiser+Lerche85}.  For these reasons, any strong spatial 
variations of the CR number density could lead to important thermal, 
chemical, and dynamical effects.

In almost all previous work it is assumed that the CR density and
spectrum do not vary significantly in space for length scales smaller
than the scale of variation in space density of CR sources or 
ionization and spallation losses \citep[e.g.][]{Hunter+97,Wolfire+03}. 
The primary rationale is that a superposition of CRs propagating 
diffusively to a point in the Galaxy from many stochastic sources 
(e.g. supernova remnants) gives an rms variation in CR density of order 
1\% for typical values of parameters (Lee 1979; see Berezinskii et al. 1990, 
sec. III.10). A more detailed calculation including the spatial distribution and 
discreteness of sources gives variations that are typically less than 10--20\%
(Busching et al. 2005).  This argument is not valid for low energy CRs 
because it neglects the possibility that self--confinement of CRs (see below) 
or ionization losses greatly reduce their propagation distance from the sources.
\nocite{Lee79,Berezinskii+90}
The evidence for CR homogeneity inferred from EGRET $\gamma$--ray data
\citep[e.g.][]{Digel+01} applies only to a spatial resolution of
approximately 5 degrees and to CR energies larger than of interest here.
Furthermore, the $\gamma$--ray emissivity is derived from an integral over
long lines of sight, to which individual clouds or cores may be 
small contributions, especially at low Galactic latitudes \citep[see][]{Aharonian01}. 

The purpose of this Letter is to show that large spatial variations
of the CR number density should indeed exist in the ISM, based on the
standard CR transport equation (\S~2), if low--energy CRs are confined along 
flux tubes by scattering from self--generated Alfv\'{e}n waves (\S~3). 
\cite{Skilling+Strong76} examined a model to exclude CRs from molecular clouds 
based on screening the CRs by ionization losses enhanced by CR self--confinement, 
but their 
model assumes that CR self--confinement sets in suddenly at the edges of molecular 
clouds because of an extremely large assumed cloud column density. The process 
examined here is completely independent of that model. Multiple magnetic mirrors 
could also lead to CR variations of a factor of a few, depending on adopted parameters  
\citep{Cesarsky+Volk78}, a process that could be more important in the presence of 
tangled fields. In reality several effects may contribute to variations on roughly 
the same scale, but the effect found here has not been previously recognized, and 
is capable of giving CR variations up to two orders of magnitude. We show that, 
in a steady state, the CR density should locally scale with the square root of the 
ion density, up to densities above which damping of the waves allows the CRs to 
stream freely (\S~3), giving a sharp decline at densities around $500$~cm$^{-3}$
(for 100~MeV protons), 
typical of the transition from diffuse gas to dark molecular clouds. These variations 
may explain the apparent discrepancy between the large ionization rate derived 
by \cite{McCall+03} for the diffuse region along the line of sight to $\zeta$ Persei 
and the ionization rate estimated in dark molecular clouds (\S~4). They should 
also result in large variations in CR pressure and heating rate in the ISM and 
elsewhere.

\nocite{LePetit+04,Liszt03}

\section{Continuity Equation for Self--Confined Cosmic Rays}

It is well--known that CRs streaming along field lines at a speed
larger than the Alfv\'{e}n speed generate MHD waves
\citep{Lerche67,Wentzel68a}. The CRs
in turn interact with these self--generated waves through resonant
pitch--angle scattering with waves whose wavenumbers are multiples
of the particle gyroradius for a particle of a given energy
\citep[see][]{Cesarsky+Kulsrud73}. As shown by \cite{Skilling71} and 
others, the waves keep the CRs confined to stream at a speed only slightly
larger than the Alfv\'{e}n speed, if the waves are not efficiently
damped. Models of Galactic CR propagation that rely on this effect
in order to increase the escape time of CRs from the Galaxy are
called self--confinement models, and are reviewed in \cite{Wentzel74}
and \cite{Cesarsky80}. Self--confinement is generally believed to be 
efficient for CRs of energies less than about 100~GeV in mostly neutral 
interstellar material \citep{Kulsrud+Cesarsky71}. 

To demonstrate how self--confinement can lead to large CR density variations,
we begin with the CR transport equation. The usual kinetic equation
for the phase space distribution, $f(\xx,p,t)$, of CRs interacting
with a background plasma or self--generated MHD wave field can be
derived from the Vlasov equation, coupled with the CR particle equation
of motion and Maxwell's equations. A number of reasonable assumptions allows 
transformation of this equation into an equation for the distribution function 
of guiding centers, assuming the CR anisotropy is small. This equation was 
derived and discussed in various forms by 
\cite{Kulsrud+Pearce69,Skilling71,Skilling75I,Earl74},
and many others. \cite{Lu+01} and \cite{Lu+02} give a useful collection of references
(see Berezinskii et al. 1990, Schlickeiser 2002 for detailed derivations).

Neglecting diffusion perpendicular to the magnetic field, introducing
a non--relativistic background medium velocity with a Galileian
transformation ($\vv \ll c$) and including continuous momentum 
losses, the result is
\begin{eqnarray}
\nonumber 
\lefteqn{
{\partial f \over \partial t}
+ \vv \cdot \nabla f
-\nabla \cdot [k(\xx,p)\nabla f]
+{p\over 3} \nabla \cdot \vv {\partial f \over {\partial p}}
=
} \\
& &
{1\over {p^2}}{\partial \over {\partial p}}
\left(
p^2 A {\partial f\over \partial p}
- p^2 {{\rm d}p\over{{\rm d}t}}f
\right)
+ S(\xx,p,t)
\label{eq0}
\end{eqnarray}
This well--known transport equation, neglecting the terms on the right, 
was derived phenomenologically by \cite{Parker65}. It is especially common
in studies of heliospheric cosmic ray transport \citep{Ferreira+Potgieter04}.

On the lhs the second term is the convection of $f$ by the 
background plasma bulk motion at velocity
$\vv$, which can include guiding center drift velocities (important in
heliospheric transport but not in most ISM conditions). When this
background motion is due to MHD waves, the appropriate velocity
depends on the distribution of the directions of wave propagation 
relative to the streaming of the CRs. If the waves are
self--generated, as we assume here, the directions of the
waves and CRs are the same, and $\vv$ can be replaced by the Alfv\'{e}n
speed, $\vv_{\rm A}$ \citep[][ch. 10]{Berezinskii+90}. The third
term represents the interaction of the CRs with the MHD waves
in the diffusion approximation; $k(\xx,p)$ is the
pitch--angle averaged spatial diffusion coefficient, which derives from 
the slight anisotropy of the CR distribution. The fourth term, 
often called the adiabatic or the Compton--Getting term, represent
momentum convection. On the rhs the first term represents momentum 
(or energy) diffusion, the second term continuous momentum losses 
due to interactions with plasma particles (e.g. spallation, ionization 
and radiation losses), and $S(\xx,p,t)$ is the source distribution function.

We can safely neglect all the terms on the rhs. It is well--known
that momentum or energy diffusion is slow compared to pitch--angle
diffusion (transformed into the spatial diffusion on the lhs), by 
a factor of order $V_A/c$ (see Berezinskii et al. 1990, ch. 10).
We neglect ionization losses because they require a column density of
order 100~g/cm$^2$ for 100~MeV protons, corresponding to large length 
scales even when tangled fields or self--confinement are taken into account. 
The sources (e.g. supernova remnants) can be assumed 
far from the relatively small ($\sim$0.01 to 10~pc) volume under consideration.
Similarly, we assume the mean magnetic field varies only over scales 
much larger than the scattering mean free path, so that we neglect mirroring 
and drifts due to mean field variations (weak focusing limit). Our 
resulting scale of variations is probably comparable to that of field 
variations, so a full calculation should include this focusing term. 

Neglecting all terms on the rhs, rearranging the 
advection term with the Compton--Getting term, and multiplying by $4\pi p^2$,
equation~(\ref{eq0}) yelds
\begin{equation}
{\partial g \over \partial t}
+\nabla \cdot (\vv g)
-\nabla \cdot [k(\xx,p)\nabla g]
={1\over 3}\nabla\cdot\vv {\partial \over {\partial p}}
\left(
p\, g
\right)
\label{eq0b}
\end{equation}
where $g(\xx,p,t) = 4\pi p^2 f(\xx,p,t)$ is the differential number 
density
of CR particles. An integration over $p$ then gives (assuming that
$p\,g$ vanishes at infinity)
\begin{equation}
{\partial n_{\rm cr} \over \partial t}
+ \nabla \cdot [\vv n_{\rm cr}
-\bar{k}(\xx)\nabla n_{\rm cr}] = 0
\label{eq1}
\end{equation}
where $n_{\rm cr}= \int_0^\infty{g(\xx,p,t)dp}$ is the total number 
density of CR particles, and $\bar{k}$ is the momentum--averaged 
spatial diffusion coefficient,
$\bar{k}(\xx)=\int_0^\infty{k(\xx,p) g(\xx,p,t)dp}/n_{\rm cr}(\xx,t)$.

Equation~(\ref{eq1}) can be cast in the form of a continuity equation by
noticing that the diffusion along the field is the divergence of a flux
that can be represented by the product of a diffusive streaming speed,
${\bf v}_{\rm diff}$, and the CR number density, $n_{\rm cr}$. We can
then define an effective CR streaming speed, ${\bf v}_{\rm st}$, as the
sum of the convection velocity, $\vv$, and the diffusive streaming 
speed, ${\bf v}_{\rm diff}$. In that case, an equation expressing a steady
state for the CR number density in an Eulerian frame is
\begin{equation}
\div({\bf v}_{\rm st}\, n_{\rm cr}) = 0
\label{eq4}
\end{equation}
A similar connection between CR transport and a continuity equation
has been noted, in different contexts, by Skilling (1971, 1975a), Earl (1974),
Schlickeiser \& Lerche (1985), Beeck \& Wibberenz (1986) and Bieber (1987).
Integrating equation~(\ref{eq4}) over a small volume, using the 
divergence theorem with a closed surface that corresponds to a segment of a 
magnetic flux tube, and using the inverse proportionality between the flux tube 
cross section and the magnetic field strength, $B$, we obtain
\begin{equation}
{v_{\rm st} \, n_{\rm cr}\over B} = {\rm const}
\label{eq5}
\end{equation}
As significant spatial variations of both $B$ and $v_{\rm st}$ are
certainly present in the ISM, we should also expect significant
variations in the CR number density, $n_{\rm cr}$,
CR pressure, $P_{\rm cr}$, and CR ionization rate, $\zeta_{\rm cr}$,
as discussed in \S~4.

\section{Cosmic Ray Streaming Velocity}

We can show how $n_{\rm cr}$ and associated quantities should vary with
ISM parameters by deriving an expression for $v_{\rm st}$.
If the CR scattering is primarily due to resonant scattering off
magnetic waves generated by the CRs themselves, $v_{\rm st}$
can be computed by requiring that the wave growth rate is balanced by 
the wave damping rate \citep[see][sec. 2.3-2.5]{Wentzel74}. 
Considering only protons, the growth rate of waves propagating in the 
direction of the magnetic field, ${\bf B}$, as a function of the CR 
streaming velocity along ${\bf B}$, $v_{\rm st}$ is \citep[eg][]{Kulsrud+Cesarsky71}:
\begin{equation}
\Gamma(k_z) = \frac{\pi(\gamma-3)\Omega_0}{4(\gamma-2)}
               \frac{n_{\rm cr}^>(k_z)}{n_{\rm i}}
               \frac{m_{\rm H}}{m_{\rm i}}
               \left( \frac{3}{\gamma}
               \frac{v_{\rm st}}{V_{\rm A}} - 1\right)
\label{eq12}
\end{equation}
\\
where $\Omega_0=eB/m_Hc$ is the non--relativistic cyclotron frequency of
a proton of mass $m_{\rm H}$, the CR spectrum has been assumed to be a
power law in momentum with exponent $\gamma$ (empirically $\gamma=4.7$
for $E\gtrsim 10$~GeV), $m_{\rm i}$ is the ion mass, and $n_{cr}^>(k_z)$
is the number density of protons with momentum $p>eB/k_zc$,
corresponding to the resonant condition.

In the mostly neutral ISM damping is primarily due to
collisions of ions with neutral particles \citep{Kulsrud+Pearce69},
because neutrals do not take part in the wave motion, as the wave
frequency is larger than the ion--neutral collision frequency at the
scales of interest.\footnote{Nonlinear cascade damping of hydromagnetic
waves \citep[e.g.][]{Skilling75II,Farmer+Goldreich04} is slow compared
to collisional damping in the cool mostly--neutral ISM.}
The ion--neutral damping rate is then:

\begin{equation}
\Gamma_{\rm in} = \frac{1}{2}n_{\rm n}\langle
                   \sigma v_{\rm n}\rangle_{\rm in}
                   \frac{m_{\rm n}}{m_{\rm i}}
\label{eq13}
\end{equation}
\\
We take the collision rate $\langle \sigma v \rangle_{\rm in}=2.1\times 
10^{-9}$~cm$^3s^{-1}$ in diffuse regions and
$\langle \sigma v \rangle_{\rm in}=1.6\times 10^{-9}$~cm$^3s^{-1}$
in molecular regions \citep{Osterbrock61}, assuming
$n({\rm He})/[n({\rm H})+2n({\rm H_2})]=0.14$.

Assuming a balance of wave growth and damping, $\Gamma=\Gamma_{\rm in}$
(such balance is reached in less than about a year for typical parameters
in the absence of nonlinear wave cascades; see Kulsrud \& Pearce 1969),
equations~(\ref{eq12}) and (\ref{eq13}) give an expression for the CR
streaming velocity \citep{Wentzel69I}:

\begin{equation}
v_{\rm st}(p)= \frac{\gamma V_{\rm A}}{3}
                \left[
                1 +
                \frac{2}{\pi} \frac{(\gamma-2)}{(\gamma-3)}
                \frac{n_{\rm n}\langle \sigma v_{\rm n}\rangle}{\Omega_0}
                \frac{n_{\rm i}}{n_{\rm cr}^>(p)}
                \frac{m_{\rm n}}{m_{\rm H}}
                \right]
\label{eq14}
\end{equation}
\begin{figure}[ht]
\centerline{
\epsfxsize=8cm \epsfbox{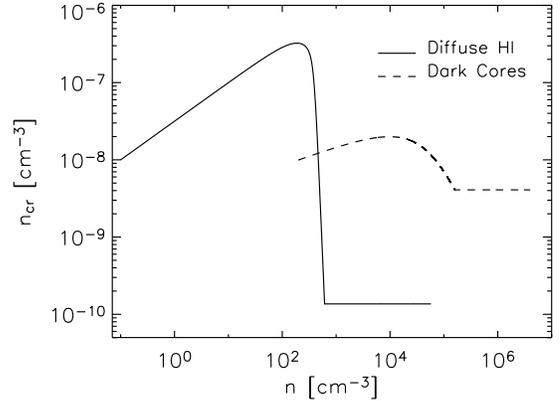}
}
\caption[]{CR density versus gas density predicted for mostly neutral 
diffuse gas and dark cloud cores by iteratively solving 
equations~(\ref{eq5}) and (\ref{eq14}). Details are given in the text.}
\label{f1}
\end{figure}
\\
The importance of the second term on the r.h.s. of
equation~(\ref{eq14}) depends on the gas density and fractional 
ionization. For mostly neutral ISM conditions and for ionizing protons of energy
$E\lesssim 100$~MeV, the second term is $\ll 1$ up to a gas density
$n_{\rm H,fs}\approx 500 (B_0/10~{\mu{\rm G}})$~cm$^{-3}$, and low energy CRs are
well confined, streaming at a velocity of order the Alfv\'{e}n 
speed. In this case, equation~(\ref{eq5}) yields:
\begin{equation}
n_{\rm cr}\propto n_{\rm i}^{1/2}
\label{eq15}
\end{equation}

Notice that the magnetic field strength has dropped out for this 
regime; it does affect the value of $n_{\rm H,fs}$ above which eq. (\ref{eq15})
no longer holds. As the gas density is increased above $n_{\rm H,fs}$, the second
term on the r.h.s. of equation~(\ref{eq14}) becomes important and the
CR streaming velocity approaches the particle velocity, causing a
drop in the value of $n_{\rm cr}$, based on equation~(\ref{eq5}).

\section{Results and Conclusions}

Figure~1 shows the result, for 100~MeV protons, of iteratively solving 
equations~(\ref{eq5}) and (\ref{eq14}) for the dependence of total CR 
number density on total gas density, taking $n_{\rm cr}^> = n_{\rm cr}$. 
We have assumed $n_{\rm i}=1.4\times 10^{-4} n$ 
\citep{Cardelli+96} and $B=B_0$ in diffuse clouds, and
$n_{\rm i}=10^{-4} (n/10^4~{\rm cm}^{-3})^{1/2} (\zeta_{\rm cr}/2\times 10^{-17}{\rm s}^{-1})^{1/2}$
\citep[e.g.][]{Elmegreen79} and $B=B_0 (n/200~{\rm cm}^{-3})^{1/2}$
\citep{Crutcher99,Bourke+01} in molecular clouds, with 
$B_0=10$~$\mu$G. These are crude approximations due to the large 
observed scatter. The vertical normalization assumes a CR energy density 
of approximately 1~eV/cm$^3$ at $n=0.1~{\rm cm}^{-3}$, corresponding to 
demodulation of the energy spectrum near the Sun \citep{Webber98}.

Corresponding variations are expected also for the CR pressure,
$P_{\rm cr} = {1\over 3} \, c \int_0^{\infty} \beta(p)\, g(\xx,p,t) \, p \, dp$
($\beta\approx 1$ for relativistic CRs, and $\beta\approx p/(mc)$
for non--relativistic CRs) and for the CR ionization rate
$\zeta_{\rm cr} = C_{\zeta}\int_0^{\infty} v(p)\, g(\xx,p,t) \, \sigma_{\rm i}(p) \, dp$
where $\sigma_{\rm i}(p)$ is the cross section for the ionization of
a hydrogen atom by a CR particle of momentum $p$, and
the factor $C_{\zeta}$ accounts for heavy CR particles and for secondary
electrons \citep[eg][]{Spitzer+Tomasko68}. 
These integrals will be computed elsewhere. Here we only stress 
that $P_{\rm cr}$ and $\zeta_{\rm cr}$ clearly increase with $n_{\rm cr}$. 
If they are nearly proportional to $n_{\rm cr}$, we then expect them to
be nearly proportional to $n_{\rm i}^{1/2}$ as well, based on
equation~(\ref{eq15}). The values of $P_{\rm cr}$ and $\zeta_{\rm cr}$
should also drop when the density is large enough to cause the
free--streaming of the CRs (when the second term on the r.h.s. of
equation~(\ref{eq14}) is large). Such drop should occur at a gas
density above $\approx 500(B_0/10~{\mu{\rm G}})$~cm$^{-3}$ in mostly neutral
diffuse regions, and above $\approx 10^5(B_0/10~{\mu{\rm G}})$~cm$^{-3}$ in dark molecular
cores, where $n_{\rm i}$ is much smaller than in diffuse regions.
The precise value of this critical density for free--streaming depends
on the magnetic field strength and on the normalization of the CR 
spectrum. Furthermore, $P_{\rm cr}$ and $\zeta_{\rm cr}$ should decrease by a 
factor of 10--50 in the transition from diffuse regions to dark molecular cores,
also due to the reduced value of $n_{\rm i}$.

Our result accounts for the heretofore unexplained enhancement of the CR ionization
rate in $\zeta$ Persei diffuse gas (McCall et al. 2003; see also Le Petit et al. 2004) 
compared to dense molecular clouds, where it is estimated using molecular abundance ratios
\citep[][and references therein]{Williams+98,Doty+02,Padoan+04} or
from the HI/H$_2$ ratio \citep[][see Liszt (2003) for a discussion of other 
evidence concerning the CR ionization rate]{Goldsmith+Li05}. Density estimates for 
the $\zeta$ Persei gas using a variety of techniques are in the range
100-400~cm$^{-3}$, putting that gas near the upper limit of our
predicted CR density for diffuse gas, just before the free--streaming
regime. In molecular regions, shielding from UV radiation allows carbon
to recombine and exist in neutral form or in molecules, resulting in 
much smaller ion density then in diffuse regions, despite the large total gas
density. This ion density is low enough that self--confinement is 
effective (up to a density of approximately $10^5$~cm$^{-3}$), but with a larger
streaming speed (larger ion Alfv\'{e}n speed) than in diffuse regions.
As a result, the CR density can be 10--50 times lower than in diffuse
regions, due to the constant--flux constraint expressed by
equation~(\ref{eq4}) (see Figure~1). Figure~1 should not be interpreted 
as predicting a one--to--one or universal relation. For example, 
we expect a large dispersion in the relation between magnetic field strength
and gas density and the CR flux normalization may vary with position relative
to nearby CR sources.

We have not discussed ionization due to electrons because the electron
spectrum is very uncertain \citep[see][]{Casadei+Bindi04}, especially 
below $100$~MeV, where it is sensitive to the models used to demodulate 
the CR electron flux at the Earth \citep[e.g.][]{Webber98} 
or used to disentangle the $\gamma$--ray bremsstrahlung, inverse compton and 
unresolved point source emission at MeV energies \citep[e.g.][]{Strong01}.
However, CR electrons, like CR protons, are confined to magnetic waves
and should therefore follow the CR protons \citep{Melrose+Wentzel70},
and have the same density variations, although the critical densities for 
free--streaming may be different.

This work points out the possibility of significant spatial variations
in the CR pressure, ionization and heating rates. However, here we 
have computed explicitly only variations of $n_{\rm cr}$, and not of $P_{\rm cr}$ and
$\zeta_{\rm cr}$. An explicit derivation of the corresponding variations
in $P_{\rm cr}$ and $\zeta_{\rm cr}$, the inclusion of electrons,
and a detailed discussion of the
observed and predicted CR ionization rates will be given elsewhere. 
The variations we find should also be important for the ionization 
fraction in planetary atmospheres and protoplanetary disk ionization 
and chemistry, as well as the heating rate and CR--initiated ion--molecule 
chemistry in molecular clouds.

\nocite{Bieber+Burger90,Bieber+87}

\acknowledgements

We are grateful to Steve Federman, Michail Malkov, Ben McCall,
Andy Strong, and Donald York for useful discussions. We thank the 
referee for many helpful comments and suggestions. J.S. acknowledges
support from NASA grants NAG5-13280 and NNG04GK43G and the NASA 
Astrobiology Institute Virtual Planetary Lead Team.


\end{document}